\documentclass[twocolumn,preprintnumbers,amssymb,amsmath,superscriptaddress,nofootinbib,groupedaddress]{revtex4}
\usepackage[dvips]{graphicx}
\usepackage{dcolumn} 
\usepackage{bm}
\usepackage{epsfig,amsmath,amssymb,verbatim,mathrsfs,array,layout,textcomp,amssymb,latexsym,slashed}
\usepackage{color}
\usepackage{hyperref}

\input epsf.tex
\newcommand{\beq}{\begin{eqnarray}}
\newcommand{\eeq}{\end{eqnarray}}

\def\ltap{\ \raise.3ex\hbox{$<$\kern-.75em\lower1ex\hbox{$\sim$}}\ }
\def\gtap{\ \raise.3ex\hbox{$>$\kern-.75em\lower1ex\hbox{$\sim$}}\ }

\def\be{\begin{equation}}
\def\ee{\end{equation}}
\def\bea{\begin{eqnarray}}
\def\eea{\end{eqnarray}}

\def\ma5{{\tt MadAnalysis5\;}}

\def\ma5{{\tt MadAnalysis5\;}}

\begin{document} 

\title{Implications of a High-Mass Diphoton Resonance for Heavy Quark Searches}

\author{Shankha Banerjee}
\affiliation{LAPTh, Universit\'e Savoie Mont Blanc, CNRS B.P. 110, F-74941 Annecy-le-Vieux, France}

\author{Daniele Barducci}
\affiliation{LAPTh, Universit\'e Savoie Mont Blanc, CNRS B.P. 110, F-74941 Annecy-le-Vieux, France}

\author{Genevi\`eve B\'elanger}
\affiliation{LAPTh, Universit\'e Savoie Mont Blanc, CNRS B.P. 110, F-74941 Annecy-le-Vieux, France}
\author{C\'edric Delaunay}
\affiliation{LAPTh, Universit\'e Savoie Mont Blanc, CNRS B.P. 110, F-74941 Annecy-le-Vieux, France}
\begin{flushright}
\preprint{\scriptsize LAPTH-033/16\vspace*{.1cm}}
\end{flushright}
\vskip .025in

\begin{abstract}
Heavy vector-like quarks coupled to a scalar $S$ will induce a coupling of this scalar to gluons and possibly (if electrically charged) photons. 
The decay of the heavy quark  into $Sq$, with $q$ being a Standard Model quark, provides, if kinematically allowed, new channels for heavy quark searches.
Inspired by naturalness considerations, we consider the case of a vector-like partner of the top quark. For illustration, we show that a singlet partner can be searched for at the 13$\,$ TeV LHC through its decay into a scalar resonance in the $2\gamma+\ell + X$ final states, especially if the diphoton branching ratio of the scalar $S$ is further enhanced by the contribution of non coloured particles.
We then show that conventional heavy quark searches are also sensitive to this new decay mode, when $S$ decays hadronically, by slightly tightening the current selection cuts. Finally, we comment about the possibility of 
disentangling, by scrutinising appropriate kinematic distributions, heavy quark decays to $St$ from other standard decay modes.
\end{abstract}

\maketitle


\section{Introduction}\label{intro}

Vector-like quarks (VLQ)  are predicted in several extensions of the standard model (SM),  from extra dimensions models to composite Higgs models and including non-minimal SUSY extensions or  grand unified theories~\cite{Graham:2009gy,Martin:2009bg,Dermisek:2012ke,Buchkremer:2012dn,Buchkremer:2013bha,Gopalakrishna:2013hua,Ellis:2014dza,Cacciapaglia:2015vrx,Low:2015qep,Franceschini:2015kwy}. 
Moreover, many of these models feature an extended scalar sector to which new fermions, in particular VLQs, can couple. A singlet in association with  VLQ has also been shown to help stabilize the electroweak vacuum~\cite{Batell:2012zw,Xiao:2014kba}.
This has two important phenomenological consequences.  First,
 charged VLQs  will  induce one-loop couplings of scalars to two photons or two gluons. Thus, the scalar $S$ produced via  gluon-fusion will have a distinctive signature at the LHC in the diphoton channel. Early hints of an excess in this channel for a resonance at 750 GeV in the first 13$\,$ TeV energy run at the Large Hadron Collider (LHC), reinforced by the reanalysis of the 8 TeV Run~\cite{1atlas750,CMS:2015dxe,ATLAS-CONF-2016-018,Aaboud:2016tru,CMS:2016owr,Khachatryan:2016hje}, have motivated a host of dedicated studies  both for the cases of the new resonance being a spin zero or two~\cite{Harigaya:2015ezk,Mambrini:2015wyu,Backovic:2015fnp,Angelescu:2015uiz,Nakai:2015ptz,Knapen:2015dap,Buttazzo:2015txu,Franceschini:2015kwy,DiChiara:2015vdm,Gupta:2015zzs,Gao:2015igz,Altmannshofer:2015xfo,
Kats:2016kuz,Strumia:2016wys,Franceschini:2016gxv}. 
  Uncoloured states were also considered as they can also lead to a large enough diphoton cross section by only increasing the diphoton partial width~\cite{Angelescu:2015uiz}, albeit with $S$ couplings dangerously approaching the non-perturbative regime~\cite{Goertz:2015nkp,Salvio:2016hnf}. 
 These hints have however not been confirmed, thus leading to an upper bound for the cross section for the production of a new scalar of   $\sigma_{pp\to S\to\gamma\gamma}<  1.5-3~ {\rm fb}$ for a narrow resonance with a mass of 750 GeV ~\cite{ATLAS-CONF-2016-059,CMS-PAS-EXO-16-027}. Second,  it opens up the possibility  for VLQs to decay into a scalar and a quark. Moreover, this VLQ decay will lead to a very clean signature if the scalar has a significant diphoton branching ratio (BR). Typically BRs of the order $10^{-3}- 10^{-2}$ are required, which are roughly in the same ball park as the diphoton BR of the SM Higgs. 
 We entertain in this letter a generic scenario where a single VLQ is coupled to a new scalar which has a non-negligible decay into  diphotons. 
 For definiteness and motivated by early hints we will consider the case where the scalar has a mass  of 750$\,$ GeV. Although similar analysis could be done for spin-2, for simplicity we consider only a scalar resonance.

Heavy VLQs are mainly pair produced via QCD interaction at the LHC, with a cross section that depends only on the VLQ mass. Decays of the VLQs are typically governed by mixings with SM quarks, yielding final states with SM quarks and electroweak (EW) gauge or Higgs bosons. 8$\,$ TeV LHC searches of VLQs with charge $2/3$ and $-1/3$ and mixing exclusively with third generation quarks excludes VLQ masses below $700-950\,$ GeV, depending on the VLQ branching ratio (BR) configuration~\cite{Khachatryan:2015oba,Khachatryan:2015gza,Aad:2015kqa}, while limits on pair-produced VLQs mixing with the light quark generations, in the $400-700\,$ GeV range, are much weaker~\cite{Aad:2015tba}\footnote{See also~\cite{Barducci:2014ila} for a phenomenological study where bounds on VLQs mixing with light generations are obtained through the recast of supersymmetry inspired searches.}. The sensitivity to high mass VLQs within the early $13\,$ TeV data is still slightly below, yet quickly reaching, that of the $8\,$ TeV run~\cite{ATLAS13-VLQ}.  
As we argue below VLQ decays into $S$ and SM quarks, whenever accessible, typically dominate over EW channels. The existence of these channels would then affect the traditional search strategies for pair-produced VLQs~\cite{Kearney:2013oia,Leskow:2014kga,Anandakrishnan:2015yfa,Serra:2015xfa} but, more importantly, their exploration would provide a complementary probe of the new physics sector associated with the new scalar  (see {\it e.g.} Ref.~\cite{Das:2015enc}).\\

In this letter, we provide a first evaluation of the impact of a heavy diphoton resonance on VLQ searches. For definiteness in the numerical studies we will assume the mass of the scalar to be $750\,$ GeV.  Motivated by the SM hierarchy problem and naturalness of the EW scale, we focus here on the possibility that the new VLQ  is mixing with third generation SM quarks. 
In particular, inspired by an approach exploited at LHC Run-1 for searching the ${\rm VLQ}\to H t$ decay~\cite{Khachatryan:2015oba}, we study the sensitivity of the LHC Run-2 to the process\footnote{Reference~\cite{Collins:2016pef} has studied in details the implications of a diphoton resonance from the decay of a singly produced vector-like quark.} $pp\to{\rm VLQ}\overline{\rm VLQ}\to S t +Y$ ($Y$ denoting any possible VLQ decay product) in the $2\gamma+\ell+X$ final states. 
 Finally, we
study the implications of the possibly large ${\rm VLQ}\to St$ decay rate on other  VLQ searches in conventional channels. We show that existing analyses could be sensitive to this decay mode, albeit with tighter kinematical selections cuts. 
\section{The model}
\label{sec:model}

We consider a simplified model that contains, in addition to the SM fields, a neutral scalar singlet $S$, and a VLQ $T,T^c$ transforming as $(\textbf{3},\textbf{1}, 2/3)$ under SU(3)$_c\times$SU(2)$_L\times$U(1)$_Y$.
The model Lagrangian is $\mathcal{L}=\mathcal{L}_{\rm SM}+\mathcal{L}_{S,T}$ where $\mathcal{L}_{\rm SM}$ is the SM Lagrangian and
\beq  
\label{eq:lag1}
 \mathcal{L}_{S,T}&=& \frac{1}{2}\left(\partial_\mu S\right)^2-\frac{m_S^2}{2}S^2 +\bar T\left(i \slashed D-M\right)T\nonumber\\
 &&- y_S S \bar T_L T_R  -  m \bar t T_L -y_T (\bar q \tilde H)T_R+{\rm h.c.}\,,
\eeq
where $H$ is the SM Higgs doublet, with $\tilde H= i\sigma_2 H^*$, and $q$ and $t$ are the SM third generation quark doublet and top quark singlet, respectively.
Both mixing parameters $m$ and $y_T v/\sqrt{2}$, with $v=(\sqrt{2}G_{\rm F})^{-1/2}\approx 246\,$ GeV, will trigger $T$ decays into SM states. Since the quark chirality of the decay products will not play any important role in our analysis, we thus consider for sake of simplicity a limit where only one mixing parameter is present and set $y_T=0$. 
In the large $M$ limit the physical top and heavy quark masses are approximately
\begin{equation}
\label{eq:topmass}
m_t^2\simeq \frac{y_t^2v^2}{2}\left(1-\frac{m^2}{M^2}\right)\,,\quad m_T^2\simeq M^2\left(1+\frac{m^2}{M^2}\right)\,,
\end{equation}
where $y_t$ is the SM top Yukawa coupling, while the mixing angles for the left and right chirality components read approximately
\begin{equation}
\begin{split}
& \tan{2\theta_R} \simeq \frac{2 m}{M}\left(1+\frac{m^2}{M^2}+\frac{ y_t^2 v^2}{2 M^2}\right)\,,\\
& \frac{\tan{2\theta_L}}{\tan{2\theta_R}} \simeq \frac{y_t v}{\sqrt{2} M}\left(1-2\frac{m^2}{M^2}\right)\,.\\
\end{split}
\end{equation}
Note that the two mixing angles are no longer related in the presence of a non-zero $\bar q\tilde H T_R$ operator.\\

In the model of Eq.~\eqref{eq:lag1} the $S-gg$ and $S- \gamma\gamma$ interactions arise dominantly through a $T$ loop\footnote{There is also a top quark loop contribution which is suppressed by the small mixing angles $s_Ls_R$ assumption (see later) and overall negligible.}, yielding the following amplitudes 
\begin{equation}
\begin{split}\label{Agg}
& \mathcal{A}_{S\to gg} = \frac{\alpha_s}{3\pi}c_{g}\delta^{ab} \left[\frac{m_S^2}{2}\left(\epsilon_1\cdot \epsilon_2\right)-(\epsilon_1\cdot p_2)(\epsilon_2\cdot p_1)\right]\,, \\
\end{split}
\end{equation}
where $p_{1,2}^\mu$ and $\epsilon_{1,2}^\mu$ are the momenta and polarization vectors of the gluons, $a,b$ are colour indices and
\beq\label{cgT}
c_g= \frac{3 y_Sc_L c_R}{4 m_T}F_{1/2}(\tau_S)\,,\quad \tau_S\equiv \frac{m_S^2}{4m_T^2}\,.
\eeq
A similar expression for $\mathcal{A}_{S\to \gamma\gamma}$ is obtained from Eq.~\eqref{Agg} through replacing $\alpha_s\to \alpha$, $\delta^{ab}\to 1$ and $c_g\to c_\gamma=6Q_T^2c_g= 8c_g/3$.
The $F_{1/2}(\tau)$ form factor, with $F_{1/2}(0)=\frac{4}{3}$, is found {\it e.g.} in Ref.~\cite{Djouadi:2005gi}.
The $S$ partial widths into $gg$, $\gamma\gamma$ and $t\bar t$ are thus
\begin{equation}
\label{eq:S-decay}
 \begin{split}
  & \Gamma_{S\to gg}=\frac{\alpha^2_s m_S^3}{72 \pi^3}\left|c_g\right|^2\,,\quad 
 \Gamma_{S\to \gamma\gamma}=\frac{\alpha^2 m_S^3}{576 \pi^3}\left|c_\gamma\right|^2,\\
  & \Gamma_{S\to t\bar t}=\frac{3 y_S^2 s_L^2 s_R^2m_S}{8 \pi}\beta_t^3,\\  
 \end{split}
\end{equation}
where $\beta_t\equiv \left(1-\frac{4 m_t^2}{m_S^2}\right)^{1/2}$. Decays into $Z\gamma$ and $ZZ$ are also predicted with, up to $m_Z/m_S$ corrections, $\Gamma_{S\to Z\gamma}/\Gamma_{S\to \gamma\gamma}\simeq2 \tan^2{\theta_W}$ and $\Gamma_{S\to ZZ}/\Gamma_{S\to \gamma\gamma}\simeq\tan^4{\theta_W}$, respectively, while $\Gamma_{S\to WW}=0$; $\theta_W$ is the weak mixing angle. Since $2 \tan^2{\theta_W}\approx0.6$ and $\tan^4{\theta_W}\approx0.1$ these decay modes will not 
significantly affect the $S$ total width, nor its diphoton BR since typically $\Gamma_{S\to\gamma\gamma}/\Gamma_{S\to gg}\simeq \mathcal{O}(\alpha^2/\alpha_s^2)\sim0.005$.\\

The mixing operators in Eq.~\eqref{eq:lag1} induces the decay of the heavy quark $T$ into $Wb$, $Zt$, $ht$ and $St$ final states, with rates
\begin{equation}
\begin{split}
& \Gamma_{T\to Wb}=\frac{\alpha s_L^2 m_T}{16\sin^2\theta_W}\lambda_{b,W}^{1/2} \zeta_{b,W}\,, \\
& \Gamma_{T\to Z t}= \frac{\alpha s_L^2 c_L^2 m_T}{32 \sin^2\theta_W\cos^2\theta_W} \lambda_{t,Z}^{1/2}   \zeta_{t,Z}\\
& \Gamma_{T\to h t}= \frac{y_t^2  m_T}{64 \pi}(c_R^2 s_L^2+c_L^2 s_R^2)  \lambda_{t,h}^{1/2}  \xi_{t,h},\\
& \Gamma_{T\to S t}= \frac{y_S^2 m_T}{32 \pi} (c_R^2 s_L^2+c_L^2 s_R^2)\lambda_{t,S}^{1/2}  \xi_{t,S}\\
\end{split}
\end{equation}
 where $\lambda_{a,b}\equiv\lambda(m_a^2/m_T^2,m_b^2/m_T^2)$ with
$\lambda(x,y)=1+x^2+y^2- 2 x -2  y -2 x y$ and similarly for $\zeta_{a,b}$ and $\xi_{a,b}$ with $\zeta(x,y)=1+x^2-2 y^2+\frac{(1-x^2)^2}{y^2}$ and
$\xi(x,y)=  1 + x^2 -y^2+ x s_{2L} s_{2R}/(c_R^2 s_L^2+c_L^2 s_R^2) $.
Note, in the limit $m_T\gg m_t$, $c_L\to 1$ and $\Gamma_{T\rightarrow Zt}\simeq\Gamma_{T\rightarrow Ht}\simeq \frac{1}{2}\Gamma_{T\rightarrow Wb}$ as expected from the equivalence theorem. In this limit, the value of BR($T\to St)$ is thus essentially determined by $y_S^2/y_t^2$.\\

The simplified model of Eq.~\eqref{eq:lag1} has three new parameters, which we choose to be $y_S$, the Yukawa coupling of the resonance with the VLQ, $m_T$ the physical VLQ mass and the RH mixing angle $\theta_R$. The loop-induced decay rates remain constant for  $\theta_R \lesssim 0.1$, while for larger mixing values the partial width in $t\bar t$ becomes comparable with the $gg$ one, leading to an overly suppressed BR in  $\gamma\gamma$. In order to ensure a large enough diphoton BR we fix $\theta_R=0.01$ for sake of definiteness.
Neglecting the $S\to t\bar t$ contribution, the diphoton BR reads
\begin{equation}
{\rm BR}(S\to\gamma\gamma) \simeq \frac{\Gamma_{S\to\gamma\gamma}}{\Gamma_{S\to gg}} = \frac{8 \alpha^2}{9 \alpha_s^2}\approx 4\times 10^{-3}.
\end{equation}
Note that the total $S$ width is dominated by the $gg$ channel and is quite small, typically sub-GeV. 

\begin{figure}[htb]
\includegraphics[width=0.45\textwidth]{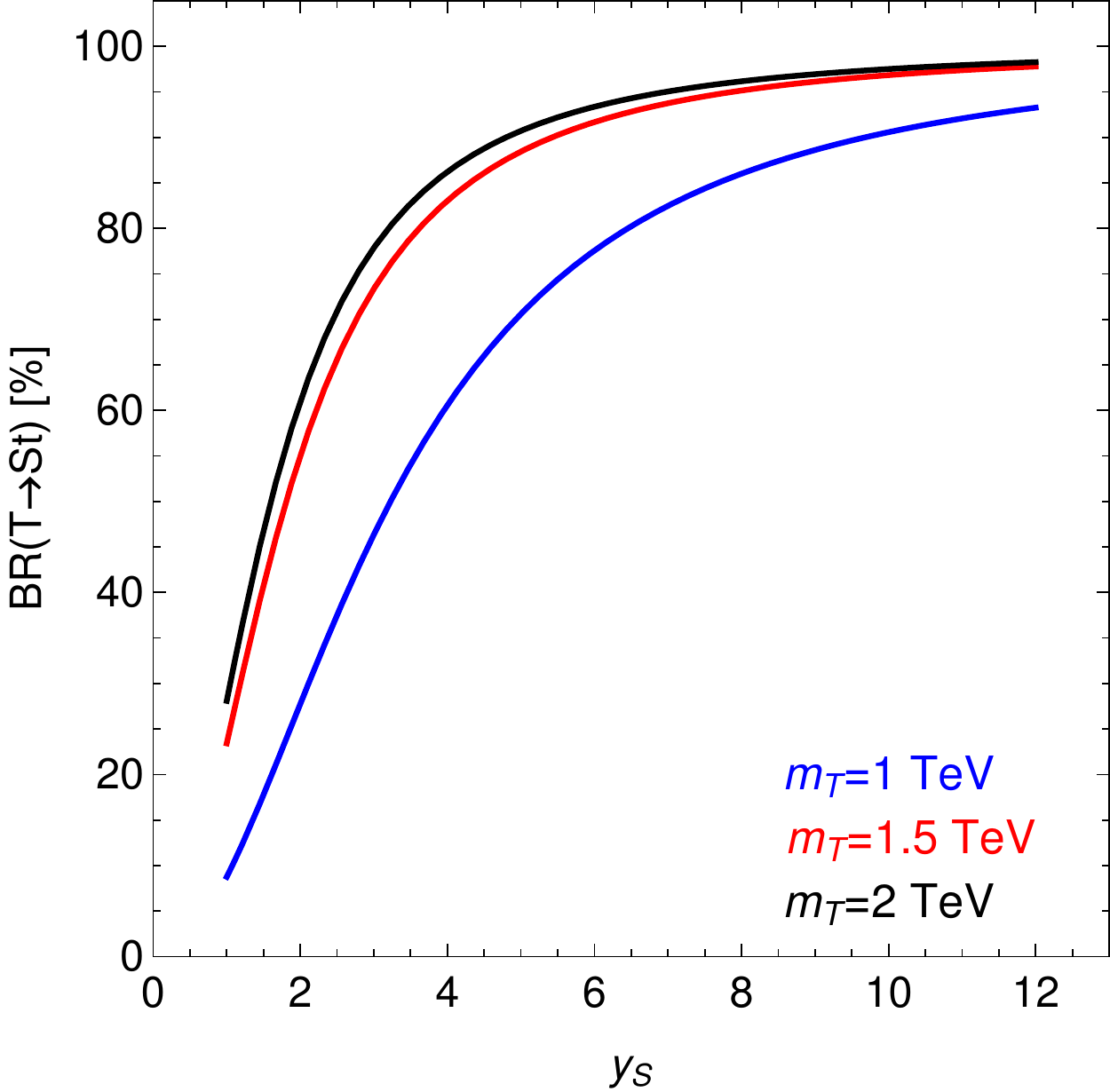}
\caption{BR$(T\to S t)$ as a function of the Yukawa coupling $y_S$ for $m_T=1\,$ TeV (blue), $1.5\,$ TeV (red) and $2\,$ TeV (black), assuming $\theta_R=0.01$.}
\label{fig:TBRs}
\end{figure}

The heavy quark BRs are almost independent of the mixing angles and, as already mentioned above, the BR into $T\to St$ is essentially set by the ratio $y_S^2/y_t^2$. This mode starts to dominate for values of $y_S\sim 4\,(2)$ for $m_Q=1\,(1.5)\,$ TeV, as shown in Fig.~\ref{fig:TBRs}.
In the model under consideration BR($S\to\gamma\gamma)$ can attain a maximum rate of $\sim0.4$\%. It is however possible to increase the $\gamma\gamma$ decay rate, without modifying the heavy quark BRs by introducing extra states, for example vector-like leptons (VLL) that will only contribute to the $S\to\gamma\gamma$ amplitude. From the expression of the loop-induced diphoton partial width
\begin{equation}
 \Gamma_{S\to \gamma\gamma}=\frac{\alpha^2 m_S^3}{576 \pi^3}| c_\gamma^T + c_\gamma^L|^2,
\end{equation}
where $c_g^T$ is given in Eq.~\eqref{cgT} and $c_\gamma^L = \sum_i \frac{3}{2} y_S^i/m_L^{i}Q_{L^i}^2 F_{1/2}(\tau_S^i)$ where the sum runs over all VLLs. Generalised to the case of heavy leptons with mass $m_L^i$, charge $Q_L^i$ and coupling $y_L^i$ to S, we can derive the ratio of the partial widths including only the heavy quark, $\Gamma_{\gamma\gamma}^T$, to the one including in  addition a heavy lepton, $\Gamma_{\gamma\gamma}^{T+L}$.
For a unit charged VLL with the same coupling $y_S$ as the heavy quark and $m_T=1000$ GeV, the ratio $\Gamma_{\gamma\gamma}^{T+L}$/$\Gamma_{\gamma\gamma}^{T}$ is $\sim$ 7 for $m_{VLL}=500$ GeV and $\sim$ 3 for $m_{VLL}=1000$ GeV. In the following, we will thus treat the $S\to\gamma\gamma$ partial width (and hence the corresponding BR) as a free parameter that can be increased by up to an order of magnitude from the value obtained using only the $T$ quark.

\section{A new channel for VLQ searches}

\label{sec:analysis}

The high $T\to S t$ BR attainable in this simplified scenario opens the possibility of searching for heavy quarks at the LHC through this unconventional decay channel, if a new scalar resonance $S$ to which the heavy quarks couple were to be discovered at the LHC. Here we study this possibility choosing for definiteness a specific value for the $S$ mass, namely $m_S\sim750$ GeV, and then commenting on how our proposed search strategy can be adapted for different $S$ mass assumptions.

During the 8 TeV run of the LHC, the CMS collaboration performed a search for pair produced heavy quarks with charge 2/3 decaying into all possible $Wb$, $Zt$ and $Ht$ combinations~\cite{Khachatryan:2015oba}. Among the various channel investigated,  an analysis was optimised for events with one heavy quark decaying into $Ht$, with the Higgs boson subsequently decaying into a $\gamma\gamma$ final state. Interestingly, despite the small SM BR$(H\to\gamma\gamma) ~\sim 2\times 10^{-3}$, this channel was exploited to set a limit of $\sim$ 600 GeV on $m_T$, under the assumption that BR$(T\to Ht)= 1$. We thus aim at exploring the sensitivity of the 13 TeV LHC for a analogous search strategy based on the $T\to S t, S\to \gamma\gamma$ decay.

The main sources of background for the 8 TeV CMS analysis are the resonant $ttH$ production, and the non resonant processes
$\gamma\gamma t\bar t$, $\gamma\gamma t$ and $\gamma\gamma+{\rm jets}$. CMS enforces selection cuts that take advantage of the resonant production of the $\gamma\gamma$ pair in the signal topology, therefore substantially reducing the non resonant background contributions. In particular  a hard cut on the leading photon, $p_T^\gamma>m_{\gamma\gamma}/2$ and an invariant mass cut on the diphoton pair, $m_{\gamma\gamma}\in [123.5,126.5]$ GeV, are applied. Moreover, high $S_T$ (defined as the scalar sum of the missing transverse momentum and the $p_T$ of the reconstructed leptons and jets) is required to fully exploit the large activity arising from the decay of pair produced heavy quarks. The events are further categorised  in fully hadronic and leptonic channels, where at least one lepton, generally arising from a top quark decay, is required.

We thus propose to adopt a similar strategy to target the $S$ resonance arising from a heavy quark decay in the leptonic channel. We adopt in particular the following event selection criteria
\begin{itemize}
\item 2 photons with $p_T>$ 10 GeV and within the kinematic acceptance of the detector,
\item $m_{\gamma\gamma} \in m_S \pm 5\%$~,
\item Leading photon with $p_T>m_{\gamma\gamma}/2$,
\item Second leading photon with $p_T>$ 25 GeV,
\item At least one lepton with $p_T>20$ GeV,
\item $S_T=E_T^{\rm miss}+H_T+\sum_{i=1}^{n_{lep}} p_T^i>770$ GeV,
\end{itemize}

where $H_T$ is the scalar sum of the transverse momentum of the reconstructed jets in the event.

 For the 8 TeV analysis the main contribution to the background in the leptonic channel comes from the non resonant processes with a yield of 0.11 events with 19.7~fb$^{-1}$ of integrated luminosity. This background is however estimated from data driven techniques, which are difficult to be accurately reproduced in a Monte Carlo simulation. Moreover, the decomposition into the $t\bar t\gamma\gamma$, $t\gamma\gamma$ and $\gamma\gamma+{\rm jets}$ component is not specified in~\cite{Khachatryan:2015oba}. 
 In order to estimate the non resonant background for the events selection proposed above we thus adopt the following strategy.
 We rescale each component of the non resonant background ($\gamma\gamma t\bar t$, $\gamma\gamma t$ and $\gamma\gamma+{\rm jets}$) by taking into account both the 
 increase in cross section due to the rescaling of the parton luminosity when going from 8 to 13 TeV of centre of mass energy as well as the difference in acceptance (at the parton level) for reconstructing a $\gamma\gamma$ invariant mass of 750 GeV, with respect to 125 GeV, with a leading photon with $p_T>m_{\gamma\gamma}/2$. We then obtain three rescaling factors, one for each non resonant background contribution, and we conservatively rescale the 8 TeV background with the highest of these factors, $\sim0.36$. With this strategy we obtain an event yield of 0.04 events with 19.7 fb $^{-1}$  at the 13 TeV LHC, while the resonant contribution, that could arise from a the $t\bar t S$ associated production, can be safely neglected due to the small $S t\bar t$ coupling that follows from the small mixing assumption. With such a small number of background events, a more refined estimate is clearly not necessary.
 Event yields for different integrated luminosity assumptions are then easily obtained as $0.04\times \mathcal L/(19.7$\;fb$^{-1}$) .

To calculate the signal cross sections for both the $gg\to S\to \gamma\gamma$ and $p p\to T\bar T$ processes, we implement the Lagrangian of Eq.~\eqref{eq:lag1} in the {\tt UFO}~\cite{Degrande:2011ua} format through the {\tt Feynrules}~\cite{Alloul:2013bka} package and use {\tt MadGraph5\_aMC@NLO}~\cite{Alwall:2014hca} as event generator. Parton showering, hadronisation and decay of unstable particles have been performed through {\tt PYTHIA v6.4}~\cite{Sjostrand:2006za} while {\tt Delphes v3.2.0}~\cite{deFavereau:2013fsa} has been employed for a fast detector simulation. Jets have been reconstructed with {\tt FastJet}~\cite{Cacciari:2011ma}, via the anti-$k_T$~\cite{Cacciari:2008gp} algorithm with cone radius 0.5, using a tuned CMS detector card suitable for performing an analysis with {\tt MadAnalysis5}~\cite{Conte:2012fm}. 

We have then generated signal samples for the processes
\begin{equation}
\begin{split}
& g g \to S \to \gamma\gamma \\
& p p \to T \bar T \to S t +  X~(+ {\rm h.c.}), S\to \gamma\gamma\\
\end{split}
\end{equation}
scanning over the following parameters values
\begin{itemize}
\item $y_S=2,4$
\item $M\in [1000$--1500] GeV
\item BR$(S\to\gamma\gamma)\in [$BR$^T_{\gamma \gamma}$, 7$\times$BR$^T_{\gamma \gamma}$]
\end{itemize}
where BR$^T_{\gamma \gamma}$ is the $S\to\gamma\gamma$ branching ratio including only  the $T$ quark. Heavy quark pair production cross sections have been normalised to the NNLO prediction computed with {\tt HATHOR}~\cite{Aliev:2010zk}, while we have applied a $k$-factor of $\sim$ 1.1 to the $\gamma\gamma$ production cross section through gluon fusion as computed by {\tt MadGraph5\_aMC@NLO}.
For this choices of parameters the LHC 8 TeV limits on resonance searches from $jj$~\cite{Aad:2014aqa,CMS-PAS-EXO-14-005}, $ZZ$~\cite{Aad:2015kna}, $Z\gamma$~\cite{Aad:2014fha} and $\gamma\gamma$~\cite{CMS:2014onr,Aad:2015mna} are fulfilled. Moreover, preliminary 13 TeV results in the $ZZ$ final state~\cite{atlas-ww} as well as limits from ATLAS searches for light dijet resonances are satisfied~\cite{ATLAS-CONF-2016-030}. 

\begin{figure}[htb]
\begin{center}
{
\includegraphics[width=0.45\textwidth]{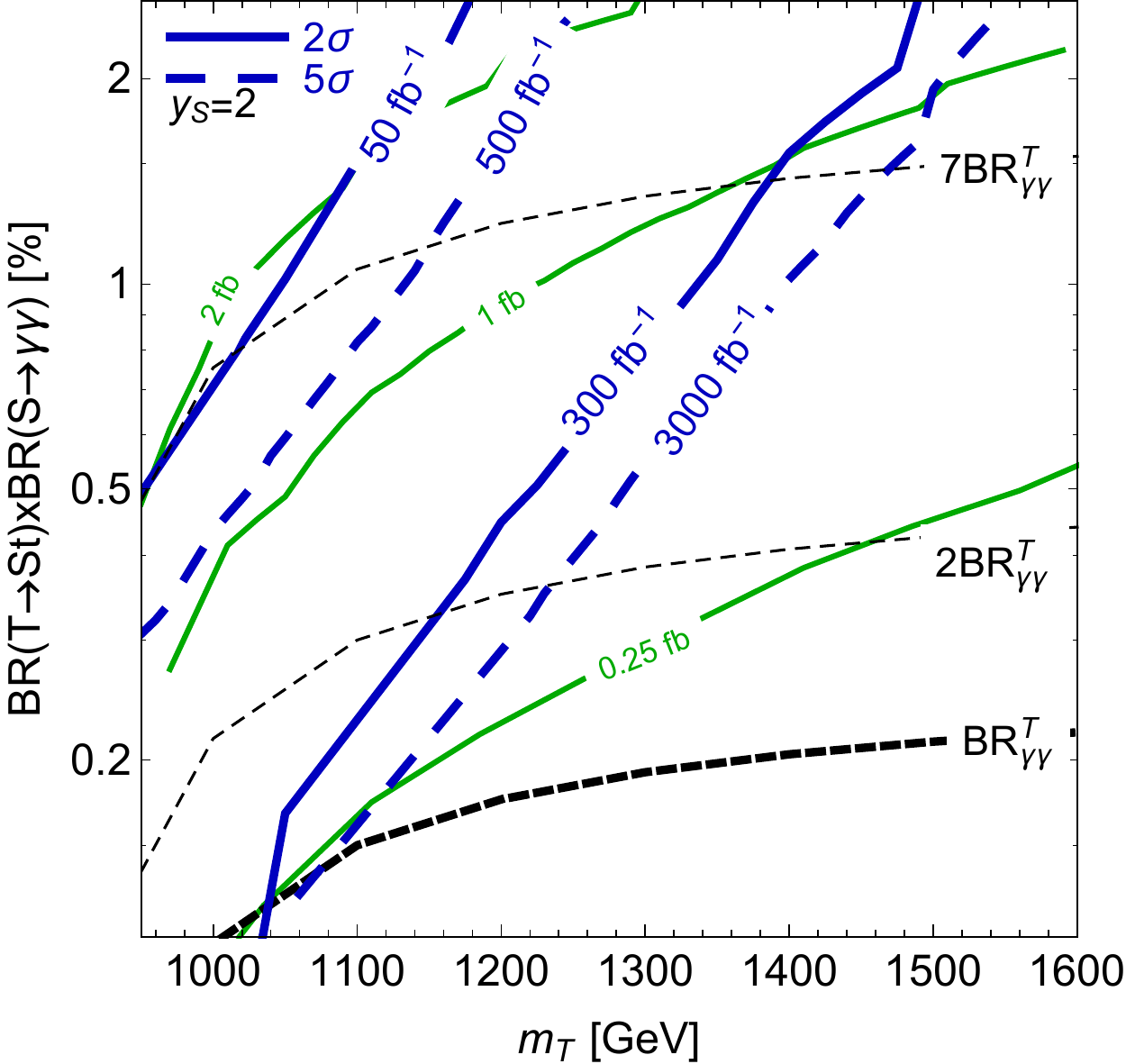}{}\hfill
\vskip 8pt
\includegraphics[width=0.45\textwidth]{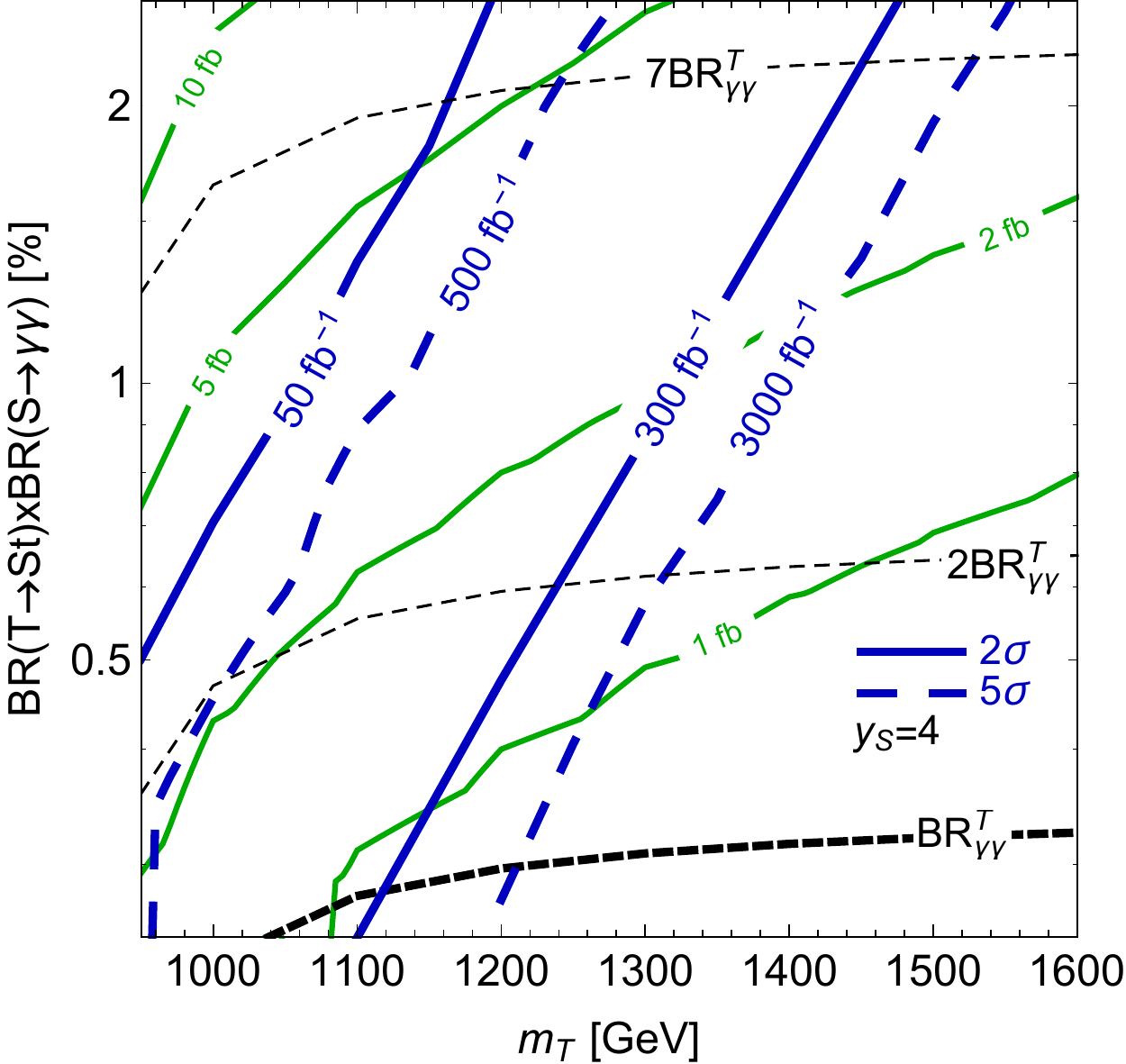}{}\hfill
\caption{2$\sigma$ and 5$\sigma$ sensitivity at the 13 TeV LHC for the $pp\to \bar T T\to 2\gamma+\ell+X$ search with various integrated luminosity options (see Figure) in the $m_T$--BR$(T\to S t)$BR$(S\to\gamma\gamma)$ plane. The black dashed lines represent various branching ratio predictions for the model, with BR$^T_{\gamma \gamma}$ the prediction with just one extra quark $T$. The green line are isocontours of constant $\sigma(gg\to\gamma\gamma)$. $y_S=2\,(4)$ is assumed in the upper (lower) panel.}
\label{fig:results}
}
\end{center}
\end{figure}

Our results for the 13 TeV reach of the LHC for VLQ  are shown in Fig.~\ref{fig:results} in the $m_T$--BR$(T\to St)\times$BR$(S\to\gamma\gamma)$ plane. The blue lines represent the 2$\sigma$ (solid) and 5$\sigma$ (dashed) isocontours, with the significance evaluated as $\alpha={\rm N}_S/\sqrt{{\rm N}_S+{\rm N}_B}$, while the green lines are isocontours of constant 
$\sigma(gg\to\gamma\gamma)$.

In the lower panel of Fig.~\ref{fig:results} we see that for $y_S=4$, and assuming BR$(S\to\gamma\gamma)=$BR$^T_{\gamma \gamma}$, up to $m_T \sim $ 1150 GeV can be probed at the $2\sigma$ level via the $T\to St$ decay channel with ${\cal L}=300 {\rm\;fb}^{-1}$.
Allowing extra particles to increase BR$(S\to\gamma\gamma)$ will give the possibility to probe higher values of $m_T$, up to $m_T\sim$ 1400 GeV with BR$(S\to\gamma\gamma)=4\times$ BR$^T_{\gamma \gamma}$.  Exploiting this channel for discovery will roughly require a factor 6 increase in  luminosity.
Note that the non observation from ATLAS and CMS of a signal excess in the diphoton mass spectrum with $\sim {\cal L}=13\;{\rm fb}^{-1}$ sets a limit of $\sim 2$ fb on $\sigma(gg\to \gamma\gamma)$~\cite{Khachatryan:2016yec,ATLAS:2016eeo}.

Similar considerations hold for smaller values of $y_s$, for example $y_S=2$ (upper panel of Fig.~\ref{fig:results}), where the constraints arising from diphoton resonant search are now less stringent and where, for BR$(S\to\gamma\gamma)=$BR$^T_{\gamma \gamma}$, the proposed search strategy attain a weaker limit.

Our analysis can be clearly adapted for different values of the $S$ mass with respect to the one analysed. 
In the case of a higher $S$ mass the SM background will be further reduced with respect to the one previously estimated, mainly due to the tighter cut on the invariant mass of the $\gamma\gamma$ system while this selection will marginally impact the signal acceptance, since the two photon will still reconstruct the $S$ invariant mass. We thus expect an overall better sensitivity for higher $S$ masses, 
which will however probe a higher minimum $m_T$ mass, $m_T>m_S+m_t$, and with a smaller $gg\to\gamma\gamma$ rate. Conversely, a smaller $S$ mass will cause a reduction of our estimated LHC reach and a higher $gg\to\gamma\gamma$ rate. In the case of a narrow $S$, this loss in sensitivity could be partially compensated by a tightening of the $\gamma\gamma$ invariant mass cut below the assumed value of 5\%.

We conclude this section by stressing that the  plane chosen for displaying our results is well suited for a general reinterpretation. In fact, $m_T$ uniquely sets the pair production rate of the heavy quark while the combination BR$(T\to St)\times$BR$(S\to\gamma\gamma)$ determines the physical cross section (before selection acceptances and decays of SM particles) in the search channel considered. The projected limits obtained in this analysis can then be applied to any model featuring a heavy quark decaying into a scalar resonance $S$ and a top quark.

\section{Impact on other VLQ searches}
\label{sec:implications}

One important question that ought to be addressed is whether and how the high $T\to S t$ rate that can be obtained in this simplified scenario will affect the standard VLQs search strategies, tailored for the direct decay of $T$ into SM final states. These searches are mainly designed to be sensitive to the processes
\begin{equation}
\begin{split}
\;& p p \to T\bar T \to W^+ W^- b \bar b \quad\\
\;& p p \to T\bar T \to Z t Z \bar t \to   W^+   W^- b \bar b Z Z\\
\;& p p \to T\bar T \to H t H \bar t \to W^+   W^- b \bar b H H\\
\end{split}
\end{equation}
in various final states~\cite{Khachatryan:2015oba}, including fully hadronic channels. While in Sec.~\ref{sec:analysis} we have exploited the similarity of the $HtH\bar t$ with the $StS\bar t$ channel in the $2\gamma+\ell+X$ final states to design an analysis sensitive to the $T\to S t$ decay mode~\footnote{Note that in case of a heavy quark with charge -1/3 decaying into $Sb$ similar search strategies can be proposed.}, we now analyse the similarities with the other two decay patterns.\\
Pair produced VLQs decaying into the heavy $S$ resonance will undergo a $T\bar T\to StS\bar t\to S S W^+ W^- b\bar b$ decay chain. Since in our scenario the $S$ is expected to decay almost 100\% of the times into a pair of gluons one finally obtains a $W^+W^- b\bar b+4j$ final state. The similarity with the final states arising from $T\to W^+ b$ and $T\to Z t$ is clearly manifest. It is thus natural to expect that standard VLQ searches could be sensitive to the $T\to S t$ decay without any significant modifications of the analysis strategies and selection cuts~\footnote{See however Ref.~\cite{Agrawal:2015dbf} for a proposal to probe the $W^+W^- b\bar b+4j$ final state with multi-jet searches designed for supersymmetric scenarios.}. 
We illustrate this for the case of an analysis targeting the $ZtZt$ final state exploited by the CMS collaboration in performing VLQ searches during LHC Run-1~\cite{Khachatryan:2015oba}. This search channel, named $OS2$, requires exactly two opposite sign leptons (electrons or muons) and at least five jets, two of which must be identified as $b$ jets, and required to have $\Delta$R$>0.3$ from the selected leptons. The invariant mass of the dilepton pair is demanded to be greater than 20 GeV and $E_T^{\rm miss}>30$ GeV is applied. 
Finally cuts on $H_T>500$ GeV and $S_T>1000$ GeV are imposed. The main background for this channel is $t\bar t+ n j$. For two benchmark points with $y_S=8$,\footnote{Note that the chosen representative value $y_S = 8$ is actually excluded by the upper bound on the diphoton cross-section. We expect however, qualitatively similar results for allowed points with $y_S = 2, 4$, with only a slight decrease in sensitivity due to the lower BR($T \rightarrow St$), see figure~\ref{fig:TBRs}.}
we check the sensitivity of the 13 TeV LHC to the $W^+W^- b\bar b+4j$ final state for a selection similar to CMS' $OS2$. We however impose tighter cuts on  $H_T>1200~{\rm GeV}$ and $S_T> 1700~{\rm GeV}$ in order to take advantage of the fact that  jets arising from the $S$ decay will  in general  be harder than the ones arising from the $Z$ ($W$) in the $T\to Zt(Wb)$ case. The number of background and signal events for the two benchmark points, corresponding to $m_T=1000$ GeV, BR$(S\to\gamma\gamma)=2\times$BR$^T_{\gamma \gamma}$ (BP$_1$) and $m_T=1100$ GeV, BR$(S\to\gamma\gamma)=$BR$^T_{\gamma \gamma}$ (BP$_2$)  are reported in Tab.~\ref{tab:nevents} for various choices of the integrated luminosity together with the value of the statistical significance $\alpha$.

\begin{table}[h!]
\centering
\begin{tabular}{c | c || c || c || c | c } 
\hline

$\mathcal{L} $ [fb$^{-1}]$~~				 &~~ N$_{\rm bkg.}$ ~~&~ N$_{{\rm BP}_1}$~ &~  $\alpha_{{\rm BP}_1}$   &~  N$_{{\rm BP}_2}$~&~  $\alpha_{{\rm BP}_2}$~~\\
\hline
\hline
100      & 439     	& 26  			& 1.2$\sigma$		& 22 		& 1.0$\sigma$ \\
300      & 1317      	& 79 			& 2.1$\sigma$ 		& 67 		& 1.8$\sigma$ \\
1000     & 4390      	& 262 			& 3.8$\sigma$ 		& 223		& 3.3$\sigma$ \\
3000     & 13170      	& 786 			& 6.7$\sigma$ 		& 674		& 5.8$\sigma$ \\
\hline
\end{tabular}
\caption{Number of background and signal events as well as the statistical significance $\alpha$ for two benchmark points with $m_T=1000$ GeV, BR$(S\to\gamma\gamma)=2\times$BR$^T_{\gamma \gamma}$ (BP$_1$) and $m_T=1100$ GeV, BR$(S\to\gamma\gamma)=$ BR$^T_{\gamma \gamma}$ (BP$_2$) after the $OS2$ selection with $H_T>1000$ GeV and $S_T>1500$ GeV for various integrated luminosity options. }
\label{tab:nevents}
\end{table}

\begin{figure}[htb]
\begin{center}
{
\includegraphics[width=0.42\textwidth]{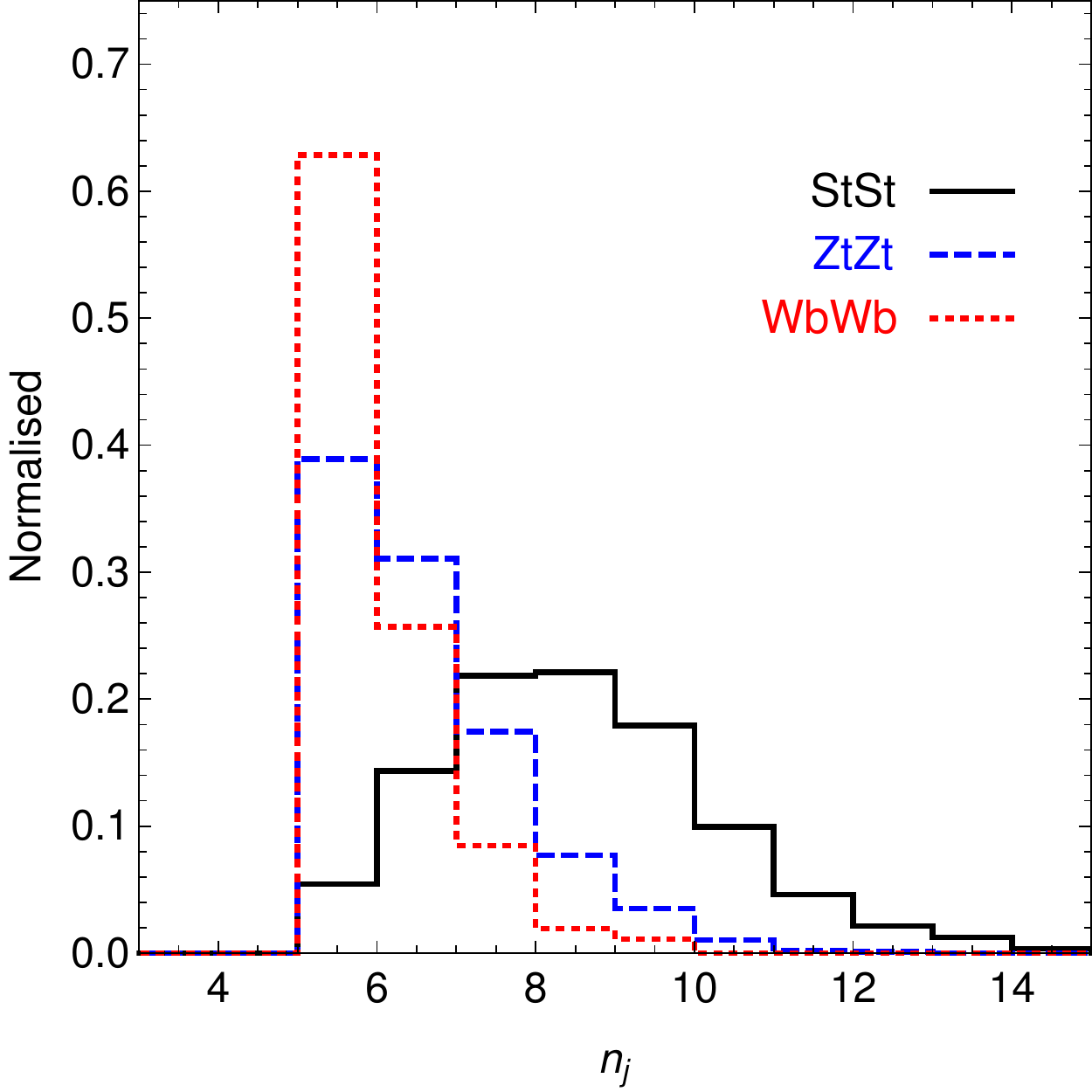}{}\hfill
\vskip 8pt
\includegraphics[width=0.45\textwidth]{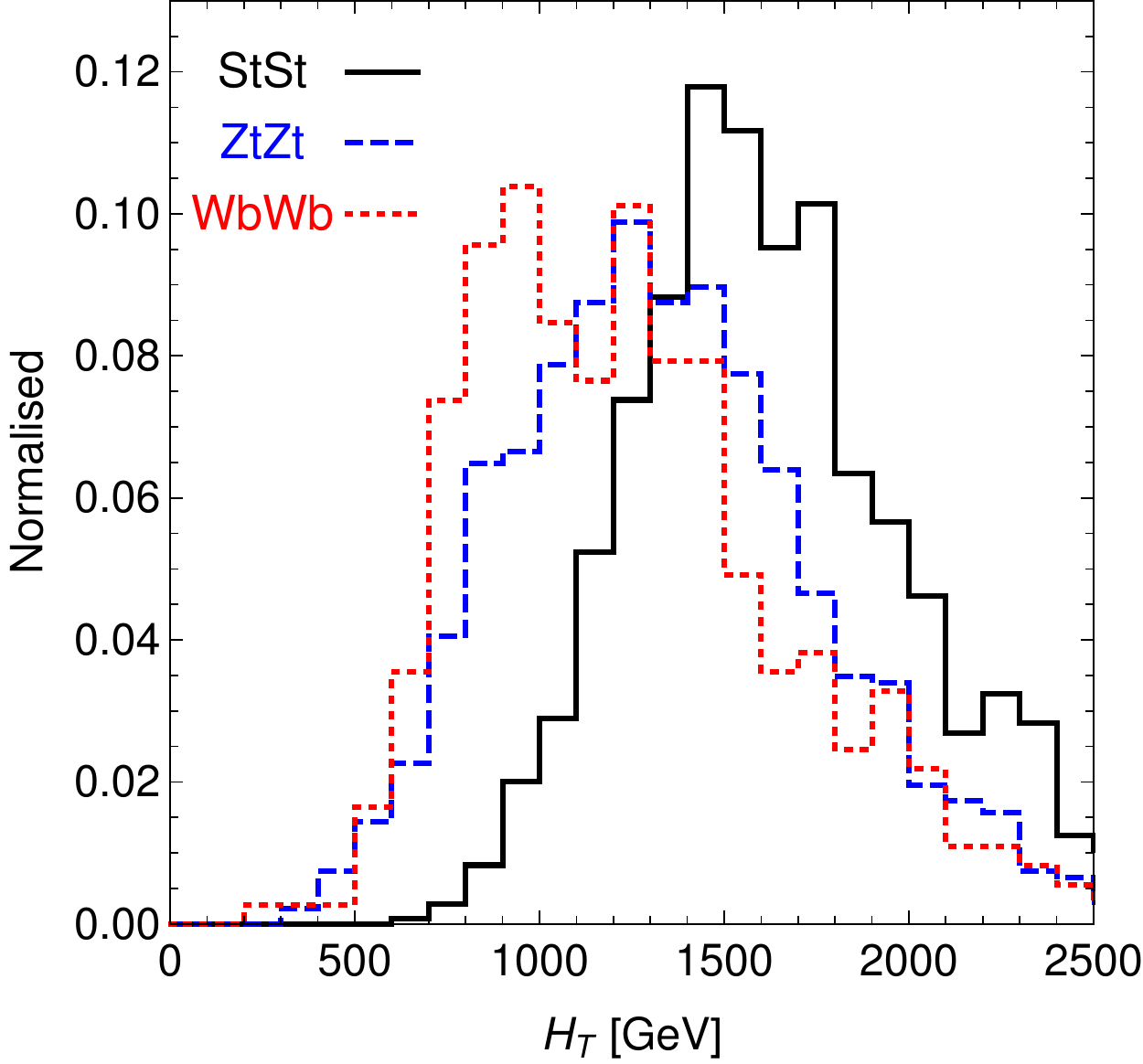}{}\hfill
\caption{Jet multiplicity and $H_T$ distributions for the case of a pair of heavy quarks with mass 1000 GeV decaying into the $W^+bW^-\bar b$, $ZtZ\bar t$ and $StS\bar t$ final states. Distributions are normalised to unity.}
\label{fig:distr}
}
\end{center}
\end{figure}

Finally, in the fortunate event that a signal is observed {\emph e.g.}, in a channel designed for the $ZtZt$ final state (as the $OS2$ previously described) it is interesting to note that it might be possible to decipher whether this arises from the decay $T\to St$ or $T\to Zt$. 
This can be achieved by exploiting the larger jet multiplicity and the higher hadronic activity found in  the $StS\bar t$ final state as compared to $W^+bW^-\bar b$, $ZtZ\bar t$ , see the corresponding distributions in Fig.~\ref{fig:distr}.

\section{Conclusions}
\label{sec:conclusions}

The decay of a heavy vector-like quark into a scalar and a light quark provide a useful search channel for VLQ at the LHC, in particular when additional particles further boost the branching ratio of the scalar into diphotons. 
Using a SU(2)$_L$ singlet heavy quark and a 750 GeV scalar as an illustration, we have shown that the 13 TeV LHC can probe VLQ decays dominantly into the top quark and the $S$ resonance in the $2\gamma+\ell+X$ channels for a large region of the parameter space compatible with upper limits on the  $pp\to S\to \gamma\gamma$ cross section.  
The projected reach we have obtained  for this channel can be directly applied to other choices of representations for VLQs since we have expressed it in terms of the heavy quark mass and the combination BR$(T\to S t)\times$BR$(S\to\gamma\gamma)$ that uniquely fixes both the $T$ pair production cross section and the event rate before kinematic selections and decay of SM particles. 
In addition, the decay of the VLQ into a scalar  resonance can also lead to a  $W^+W^-b\bar b +4j$ final state. We have shown that current VLQ searches tailored for the direct decay of the heavy quark into SM states can be sensitive to this final state without a drastic modification of the selection cuts. 
Although we have focused our analysis on the specific case of a 750 GeV scalar, our results can easily be extended to any high mass state  lighter than the vector-like quark and decaying into diphotons.
Finally, in the event  an excess is observed in an analysis targeting pair produced heavy quarks we observe that various kinematic distributions can be exploited to disentangle whether this excess arises from the $T\to St$ decay or from standard heavy quark decay modes.

\section*{Acknowledgements}
We wish to thank Fawzi Boudjema, Devdatta Majumder and Emanuele Re for useful discussions. This work is supported by the ``Investissements d'avenir, Labex ENIGMASS'', by the French ANR,  Project DMAstro-LHC, ANR-12-BS05-006, by the Research Executive Agency (REA) of the European Union under the Grant Agreement PITN-GA-2012-316704 (``HiggsTools") and by the Indo French LIA THEP (Theoretical High Energy Physics) of the CNRS.\\

\bibliographystyle{ieeetr}
\bibliography{refs}

\begin{thebibliography}{10}

\bibitem{Graham:2009gy}
P.~W. Graham, A.~Ismail, S.~Rajendran, and P.~Saraswat, ``{A Little Solution to
  the Little Hierarchy Problem: A Vector-like Generation},'' {\em Phys. Rev.},
  vol.~D81, p.~055016, 2010.

\bibitem{Martin:2009bg}
S.~P. Martin, ``{Extra vector-like matter and the lightest Higgs scalar boson
  mass in low-energy supersymmetry},'' {\em Phys. Rev.}, vol.~D81, p.~035004,
  2010.

\bibitem{Dermisek:2012ke}
R.~Dermisek, ``{Unification of gauge couplings in the standard model with extra
  vectorlike families},'' {\em Phys. Rev.}, vol.~D87, no.~5, p.~055008, 2013.

\bibitem{Buchkremer:2012dn}
M.~Buchkremer and A.~Schmidt, ``{Long-lived heavy quarks : a review},'' {\em
  Adv. High Energy Phys.}, vol.~2013, p.~690254, 2013.

\bibitem{Buchkremer:2013bha}
M.~Buchkremer, G.~Cacciapaglia, A.~Deandrea, and L.~Panizzi, ``{Model
  Independent Framework for Searches of Top Partners},'' {\em Nucl. Phys.},
  vol.~B876, pp.~376--417, 2013.

\bibitem{Gopalakrishna:2013hua}
S.~Gopalakrishna, T.~Mandal, S.~Mitra, and G.~Moreau, ``{LHC Signatures of
  Warped-space Vectorlike Quarks},'' {\em JHEP}, vol.~08, p.~079, 2014.

\bibitem{Ellis:2014dza}
S.~A.~R. Ellis, R.~M. Godbole, S.~Gopalakrishna, and J.~D. Wells, ``{Survey of
  vector-like fermion extensions of the Standard Model and their
  phenomenological implications},'' {\em JHEP}, vol.~09, p.~130, 2014.

\bibitem{Cacciapaglia:2015vrx}
G.~Cacciapaglia and A.~Parolini, ``{Light ?t Hooft top partners},'' {\em Phys.
  Rev.}, vol.~D93, no.~7, p.~071701, 2016.

\bibitem{Low:2015qep}
M.~Low, A.~Tesi, and L.-T. Wang, ``{A pseudoscalar decaying to photon pairs in
  the early LHC Run 2 data},'' {\em JHEP}, vol.~03, p.~108, 2016.

\bibitem{Franceschini:2015kwy}
R.~Franceschini, G.~F. Giudice, J.~F. Kamenik, M.~McCullough, A.~Pomarol,
  R.~Rattazzi, M.~Redi, F.~Riva, A.~Strumia, and R.~Torre, ``{What is the
  $\gamma \gamma$ resonance at 750 GeV?},'' {\em JHEP}, vol.~03, p.~144, 2016.

\bibitem{Batell:2012zw}
B.~Batell, S.~Jung, and H.~M. Lee, ``{Singlet Assisted Vacuum Stability and the
  Higgs to Diphoton Rate},'' {\em JHEP}, vol.~01, p.~135, 2013.

\bibitem{Xiao:2014kba}
M.-L. Xiao and J.-H. Yu, ``{Stabilizing electroweak vacuum in a vectorlike
  fermion model},'' {\em Phys. Rev.}, vol.~D90, no.~1, p.~014007, 2014.
\newblock [Addendum: Phys. Rev.D90,no.1,019901(2014)].

\bibitem{1atlas750}
``{Search for resonances decaying to photon pairs in 3.2 fb$^{-1}$ of $pp$
  collisions at $\sqrt{s}$ = 13 TeV with the ATLAS detector},'' {\em
  \href{http://cds.cern.ch/record/2114853/files/ATLAS-CONF-2015-081.pdf}{ATLAS-CONF-2015-081}},
  2015.

\bibitem{CMS:2015dxe}
``{Search for new physics in high mass diphoton events in proton-proto
  collisions at 13TeV},'' {\em
  \href{http://cds.cern.ch/record/2114808/files/EXO-15-004-pas.pdf}{CMS-PAS-EXO-15-004}},
  2015.

\bibitem{ATLAS-CONF-2016-018}
``{Search for resonances in diphoton events with the ATLAS detector at
  $\sqrt{s}$ = 13 TeV},'' {\em
  \href{http://cds.cern.ch/record/2141568/files/ATLAS-CONF-2016-018.pdf}{ATLAS-CONF-2016-018}},
  2016.

\bibitem{Aaboud:2016tru}
M.~Aaboud {\em et~al.}, ``{Search for resonances in diphoton events at
  $\sqrt{s}$=13 TeV with the ATLAS detector},'' {\em JHEP}, vol.~09, p.~001,
  2016.

\bibitem{CMS:2016owr}
``{Search for new physics in high mass diphoton events in
  $3.3~\mathrm{fb}^{-1}$ of proton-proton collisions at
  $\sqrt{s}=13~\mathrm{TeV}$ and combined interpretation of searches at
  $8~\mathrm{TeV}$ and $13~\mathrm{TeV}$},'' {\em
  \href{http://cds.cern.ch/record/2139899/files/EXO-16-018-pas.pdf}{CMS-PAS-EXO-16-018}},
  2016.

\bibitem{Khachatryan:2016hje}
V.~Khachatryan {\em et~al.}, ``{Search for Resonant Production of High-Mass
  Photon Pairs in Proton-Proton Collisions at $\sqrt s$ =8 and 13 TeV},'' {\em
  Phys. Rev. Lett.}, vol.~117, no.~5, p.~051802, 2016.

\bibitem{Harigaya:2015ezk}
K.~Harigaya and Y.~Nomura, ``{Composite Models for the 750 GeV Diphoton
  Excess},'' {\em Phys. Lett.}, vol.~B754, pp.~151--156, 2016.

\bibitem{Mambrini:2015wyu}
Y.~Mambrini, G.~Arcadi, and A.~Djouadi, ``{The LHC diphoton resonance and dark
  matter},'' {\em Phys. Lett.}, vol.~B755, pp.~426--432, 2016.

\bibitem{Backovic:2015fnp}
M.~Backovic, A.~Mariotti, and D.~Redigolo, ``{Di-photon excess illuminates Dark
  Matter},'' {\em JHEP}, vol.~03, p.~157, 2016.

\bibitem{Angelescu:2015uiz}
A.~Angelescu, A.~Djouadi, and G.~Moreau, ``{Scenarii for interpretations of the
  LHC diphoton excess: two Higgs doublets and vector-like quarks and
  leptons},'' {\em Phys. Lett.}, vol.~B756, pp.~126--132, 2016.

\bibitem{Nakai:2015ptz}
Y.~Nakai, R.~Sato, and K.~Tobioka, ``{Footprints of New Strong Dynamics via
  Anomaly and the 750 GeV Diphoton},'' {\em Phys. Rev. Lett.}, vol.~116,
  no.~15, p.~151802, 2016.

\bibitem{Knapen:2015dap}
S.~Knapen, T.~Melia, M.~Papucci, and K.~Zurek, ``{Rays of light from the
  LHC},'' {\em Phys. Rev.}, vol.~D93, no.~7, p.~075020, 2016.

\bibitem{Buttazzo:2015txu}
D.~Buttazzo, A.~Greljo, and D.~Marzocca, ``{Knocking on new physics? door with
  a scalar resonance},'' {\em Eur. Phys. J.}, vol.~C76, no.~3, p.~116, 2016.

\bibitem{DiChiara:2015vdm}
S.~Di~Chiara, L.~Marzola, and M.~Raidal, ``{First interpretation of the 750 GeV
  diphoton resonance at the LHC},'' {\em Phys. Rev.}, vol.~D93, no.~9,
  p.~095018, 2016.

\bibitem{Gupta:2015zzs}
R.~S. Gupta, S.~Jäger, Y.~Kats, G.~Perez, and E.~Stamou, ``{Interpreting a 750
  GeV Diphoton Resonance},'' {\em JHEP}, vol.~07, p.~145, 2016.

\bibitem{Gao:2015igz}
J.~Gao, H.~Zhang, and H.~X. Zhu, ``{Diphoton excess at 750 GeV: gluon?gluon
  fusion or quark?antiquark annihilation?},'' {\em Eur. Phys. J.}, vol.~C76,
  no.~6, p.~348, 2016.

\bibitem{Altmannshofer:2015xfo}
W.~Altmannshofer, J.~Galloway, S.~Gori, A.~L. Kagan, A.~Martin, and J.~Zupan,
  ``{750 GeV diphoton excess},'' {\em Phys. Rev.}, vol.~D93, no.~9, p.~095015,
  2016.

\bibitem{Kats:2016kuz}
Y.~Kats and M.~J. Strassler, ``{Resonances from QCD bound states and the 750
  GeV diphoton excess},'' {\em JHEP}, vol.~05, p.~092, 2016.

\bibitem{Strumia:2016wys}
A.~Strumia, ``{Interpreting the 750 GeV digamma excess: a review},'' 2016.

\bibitem{Franceschini:2016gxv}
R.~Franceschini, G.~F. Giudice, J.~F. Kamenik, M.~McCullough, F.~Riva,
  A.~Strumia, and R.~Torre, ``{Digamma, what next?},'' {\em JHEP}, vol.~07,
  p.~150, 2016.

\bibitem{Goertz:2015nkp}
F.~Goertz, J.~F. Kamenik, A.~Katz, and M.~Nardecchia, ``{Indirect Constraints
  on the Scalar Di-Photon Resonance at the LHC},'' {\em JHEP}, vol.~05, p.~187,
  2016.

\bibitem{Salvio:2016hnf}
A.~Salvio, F.~Staub, A.~Strumia, and A.~Urbano, ``{On the maximal diphoton
  width},'' {\em JHEP}, vol.~03, p.~214, 2016.

\bibitem{ATLAS-CONF-2016-059}
``{Search for scalar diphoton resonances with 15.4~fb$^{-1}$ of data collected
  at $\sqrt{s}$=13 TeV in 2015 and 2016 with the ATLAS detector},'' Tech. Rep.
  ATLAS-CONF-2016-059, CERN, Geneva, Aug 2016.

\bibitem{CMS-PAS-EXO-16-027}
``{Search for resonant production of high mass photon pairs using
  $12.9\,\mathrm{fb^{-1}}$ of proton-proton collisions at $\sqrt{s} =
  13~\mathrm{TeV}$ and combined interpretation of searches at 8 and 13 TeV},''
  Tech. Rep. CMS-PAS-EXO-16-027, CERN, Geneva, 2016.

\bibitem{Khachatryan:2015oba}
V.~Khachatryan {\em et~al.}, ``{Search for vector-like charge 2/3 T quarks in
  proton-proton collisions at sqrt(s) = 8 TeV},'' {\em Phys. Rev.}, vol.~D93,
  no.~1, p.~012003, 2016.

\bibitem{Khachatryan:2015gza}
V.~Khachatryan {\em et~al.}, ``{Search for pair-produced vector-like B quarks
  in proton-proton collisions at $\sqrt{s}$ = 8 TeV},'' 2015.

\bibitem{Aad:2015kqa}
G.~Aad {\em et~al.}, ``{Search for production of vector-like quark pairs and of
  four top quarks in the lepton-plus-jets final state in $pp$ collisions at
  $\sqrt{s}=8$ TeV with the ATLAS detector},'' {\em JHEP}, vol.~08, p.~105,
  2015.

\bibitem{Aad:2015tba}
G.~Aad {\em et~al.}, ``{Search for pair production of a new heavy quark that
  decays into a $W$ boson and a light quark in $pp$ collisions at $\sqrt{s} =
  8$ TeV with the ATLAS detector},'' {\em Phys. Rev.}, vol.~D92, no.~11,
  p.~112007, 2015.

\bibitem{Barducci:2014ila}
D.~Barducci, A.~Belyaev, M.~Buchkremer, G.~Cacciapaglia, A.~Deandrea,
  S.~De~Curtis, J.~Marrouche, S.~Moretti, and L.~Panizzi, ``{Framework for
  Model Independent Analyses of Multiple Extra Quark Scenarios},'' {\em JHEP},
  vol.~12, p.~080, 2014.

\bibitem{ATLAS13-VLQ}
``{Search for production of vector-like top quark pairs and of four top quarks
  in the lepton-plus-jets final state in $pp$ collisions at $\sqrt{s}=13$ TeV
  with the ATLAS detector},'' {\em
  \href{http://cds.cern.ch/record/2140998/files/ATLAS-CONF-2016-013.pdf}{ATLAS-CONF-2016-013}},
  2016.

\bibitem{Kearney:2013oia}
J.~Kearney, A.~Pierce, and J.~Thaler, ``{Top Partner Probes of Extended Higgs
  Sectors},'' {\em JHEP}, vol.~08, p.~130, 2013.

\bibitem{Leskow:2014kga}
E.~Coluccio~Leskow, T.~A.~W. Martin, and A.~de~la Puente, ``{Vector-like quarks
  with a scalar triplet},'' {\em Phys. Lett.}, vol.~B743, pp.~366--376, 2015.

\bibitem{Anandakrishnan:2015yfa}
A.~Anandakrishnan, J.~H. Collins, M.~Farina, E.~Kuflik, and M.~Perelstein,
  ``{Odd Top Partners at the LHC},'' {\em Phys. Rev.}, vol.~D93, no.~7,
  p.~075009, 2016.

\bibitem{Serra:2015xfa}
J.~Serra, ``{Beyond the Minimal Top Partner Decay},'' {\em JHEP}, vol.~09,
  p.~176, 2015.

\bibitem{Das:2015enc}
K.~Das and S.~K. Rai, ``{750 GeV diphoton excess in a U(1) hidden symmetry
  model},'' {\em Phys. Rev.}, vol.~D93, no.~9, p.~095007, 2016.

\bibitem{Collins:2016pef}
J.~H. Collins, C.~Csaki, J.~A. Dror, and S.~Lombardo, ``{Novel kinematics from
  a custodially protected diphoton resonance},'' {\em Phys. Rev.}, vol.~D93,
  no.~11, p.~115001, 2016.

\bibitem{Djouadi:2005gi}
A.~Djouadi, ``{The Anatomy of electro-weak symmetry breaking. I: The Higgs
  boson in the standard model},'' {\em Phys. Rept.}, vol.~457, pp.~1--216,
  2008.

\bibitem{Degrande:2011ua}
C.~Degrande, C.~Duhr, B.~Fuks, D.~Grellscheid, O.~Mattelaer, and T.~Reiter,
  ``{UFO - The Universal FeynRules Output},'' {\em Comput. Phys. Commun.},
  vol.~183, pp.~1201--1214, 2012.

\bibitem{Alloul:2013bka}
A.~Alloul, N.~D. Christensen, C.~Degrande, C.~Duhr, and B.~Fuks, ``{FeynRules
  2.0 - A complete toolbox for tree-level phenomenology},'' {\em Comput. Phys.
  Commun.}, vol.~185, pp.~2250--2300, 2014.

\bibitem{Alwall:2014hca}
J.~Alwall, R.~Frederix, S.~Frixione, V.~Hirschi, F.~Maltoni, O.~Mattelaer,
  H.~S. Shao, T.~Stelzer, P.~Torrielli, and M.~Zaro, ``{The automated
  computation of tree-level and next-to-leading order differential cross
  sections, and their matching to parton shower simulations},'' {\em JHEP},
  vol.~07, p.~079, 2014.

\bibitem{Sjostrand:2006za}
T.~Sjostrand, S.~Mrenna, and P.~Z. Skands, ``{PYTHIA 6.4 Physics and Manual},''
  {\em JHEP}, vol.~05, p.~026, 2006.

\bibitem{deFavereau:2013fsa}
J.~de~Favereau, C.~Delaere, P.~Demin, A.~Giammanco, V.~Lemaître, A.~Mertens,
  and M.~Selvaggi, ``{DELPHES 3, A modular framework for fast simulation of a
  generic collider experiment},'' {\em JHEP}, vol.~02, p.~057, 2014.

\bibitem{Cacciari:2011ma}
M.~Cacciari, G.~P. Salam, and G.~Soyez, ``{FastJet User Manual},'' {\em Eur.
  Phys. J.}, vol.~C72, p.~1896, 2012.

\bibitem{Cacciari:2008gp}
M.~Cacciari, G.~P. Salam, and G.~Soyez, ``{The Anti-k(t) jet clustering
  algorithm},'' {\em JHEP}, vol.~04, p.~063, 2008.

\bibitem{Conte:2012fm}
E.~Conte, B.~Fuks, and G.~Serret, ``{MadAnalysis 5, A User-Friendly Framework
  for Collider Phenomenology},'' {\em Comput. Phys. Commun.}, vol.~184,
  pp.~222--256, 2013.

\bibitem{Aliev:2010zk}
M.~Aliev, H.~Lacker, U.~Langenfeld, S.~Moch, P.~Uwer, and M.~Wiedermann,
  ``{HATHOR: HAdronic Top and Heavy quarks crOss section calculatoR},'' {\em
  Comput. Phys. Commun.}, vol.~182, pp.~1034--1046, 2011.

\bibitem{Aad:2014aqa}
G.~Aad {\em et~al.}, ``{Search for new phenomena in the dijet mass distribution
  using $p-p$ collision data at $\sqrt{s}=8$ TeV with the ATLAS detector},''
  {\em Phys. Rev.}, vol.~D91, no.~5, p.~052007, 2015.

\bibitem{CMS-PAS-EXO-14-005}
``{Search for Resonances Decaying to Dijet Final States at $\sqrt{s} = 8$ TeV
  with Scouting Data},'' {\em
  \href{https://cds.cern.ch/record/2063491}{CMS-PAS-EXO-14-005}}, 2014.

\bibitem{Aad:2015kna}
G.~Aad {\em et~al.}, ``{Search for an additional, heavy Higgs boson in the
  $H\rightarrow ZZ$ decay channel at $\sqrt{s} = 8\;\text{ TeV }$ in $pp$
  collision data with the ATLAS detector},'' {\em Eur. Phys. J.}, vol.~C76,
  no.~1, p.~45, 2016.

\bibitem{Aad:2014fha}
G.~Aad {\em et~al.}, ``{Search for new resonances in $W\gamma$ and $Z\gamma$
  final states in $pp$ collisions at $\sqrt s=8$ TeV with the ATLAS
  detector},'' {\em Phys. Lett.}, vol.~B738, pp.~428--447, 2014.

\bibitem{CMS:2014onr}
``{Search for an Higgs Like resonance in the diphoton mass spectra above 150
  GeV with 8 TeV data},'' {\em
  \href{http://cds.cern.ch/record/1714076/files/HIG-14-006-pas.pdf}{CMS-PAS-HIG-14-006}},
  2014.

\bibitem{Aad:2015mna}
G.~Aad {\em et~al.}, ``{Search for high-mass diphoton resonances in $pp$
  collisions at $\sqrt{s}=8$ TeV with the ATLAS detector},'' {\em Phys. Rev.},
  vol.~D92, no.~3, p.~032004, 2015.

\bibitem{atlas-ww}
``{Search for diboson resonances in the llqq final state in pp collisions at
  $\sqrt{s}$ = 13 TeV with the ATLAS detector},'' {\em
  \href{http://cds.cern.ch/record/2114843/files/ATLAS-CONF-2015-071.pdf}{ATLAS-CONF-2015-071}},
  2015.

\bibitem{ATLAS-CONF-2016-030}
``{Search for light dijet resonances with the ATLAS detector using a
  Trigger-Level Analysis in LHC pp collisions at $\sqrt{s}=13$~TeV},'' Tech.
  Rep. ATLAS-CONF-2016-030, CERN, Geneva, Jun 2016.

\bibitem{Khachatryan:2016yec}
V.~Khachatryan {\em et~al.}, ``{Search for high-mass diphoton resonances in
  proton-proton collisions at 13 TeV and combination with 8 TeV search},''
  2016.

\bibitem{ATLAS:2016eeo}
T.~A. collaboration, ``{Search for scalar diphoton resonances with
  15.4~fb$^{-1}$ of data collected at $\sqrt{s}$=13 TeV in 2015 and 2016 with
  the ATLAS detector},'' 2016.

\bibitem{Agrawal:2015dbf}
P.~Agrawal, J.~Fan, B.~Heidenreich, M.~Reece, and M.~Strassler, ``{Experimental
  Considerations Motivated by the Diphoton Excess at the LHC},'' {\em JHEP},
  vol.~06, p.~082, 2016.

\end{thebibliography}

\end{document}